%% file: paper.tex
\pdfoutput=1 
\documentclass[a4paper]{article}

\usepackage{spconf,amsmath,graphicx,bm}
\usepackage{color}

\def\vec#1{\ensuremath{\bm{{#1}}}}

\title{DEEP CONTEXT: END-TO-END CONTEXTUAL SPEECH RECOGNITION}
\name{Golan Pundak, Tara N. Sainath, Rohit Prabhavalkar, Anjuli Kannan, Ding Zhao}
\address{
  Google Inc., USA
}

\begin{document}

\maketitle

\input{abstract}


\input{intro}

\section{Background}

\input{las}

\input{standard_context}

\input{model}

\input{experiments}

\input{results}

\input{conclusion}

\section{Acknowledgements}

We would like to thank Ian Williams, Petar Aleksic, Assaf Hurwitz Michaely, Uri Alon,
Justin Scheiner, Yanzhang (Ryan) He, Deepti Bhatia, Johan Schalkwyk and Michiel Bacchiani for their help and useful comments.

\bibliographystyle{IEEEbib}

\bibliography{mybib}

\end{document}

%% file: abstract.tex
\begin{abstract}

In automatic speech recognition (ASR) what a user says depends on the particular
context she is in.
Typically, this context is represented as a set of word n-grams.
In this work, we present a novel, all-neural, end-to-end (E2E) ASR system that utilizes such context.
Our approach, which we refer to as Contextual Listen, Attend and Spell (CLAS) jointly-optimizes the ASR components
along with embeddings of the context n-grams.
During inference, the CLAS system can be presented with context phrases
which might contain out-of-vocabulary (OOV) terms not seen during training.
% In automatic speech recognition (ASR) what a user says depends on the particular
% context she is in. Typically, the context information is represented as a set of n-grams,
% which are constructed independetly from other components of the ASR pipeline, and incorporated
% as an additional language model (LM) component during decoding. The most straightforward approach
% to incorporating contextual information into sequence-to-sequence models is via shallow fusion, namely interpolating
% the scores of the sequence-to-sequence and contextual LM during decoding. However, this shallow fusion approach reduces
% a large benefit of sequence-to-sequence models, namely the joint optimization of all components of the ASR pipeline in one neural network.
%
% In this paper, we address ths issue by developing an all-neural approach for injecting
% contextual information into a sequence-to-sequence model, which we refer to as Contextual Listen, Attend and Spell (CLAS).
We compare our proposed system to a more traditional contextualization approach,
which performs shallow-fusion between independently trained LAS and contextual n-gram models during beam search.
Across a number of tasks, we find that the proposed CLAS system outperforms the
baseline method by as much as 68\% relative WER, indicating the advantage of
joint optimization over individually trained components.
%. Specifically,

%examine injecting deep embeddings of contextual information
%into the ``Listen, Attend and Spell" (LAS) sequence-to-sequence model.
%We develop a novel neural component that attends to embeddings of the context,
%and train it jointly with the rest of the LAS model.
%The proposed model, which we refer to as Contextual-LAS (CLAS), is
%in sharp contrast to traditional contextualization methods that only modify the
%model's output probabilites.

%A key aspect of our model is its ability to incorporate out-of-vocabulary (OOV)
%terms as contextual phrases during inference.

%TS: we need to stress that everything is jointly optimized in the intro
\end{abstract}

%% file: intro.tex
\section{Introduction}
As speech technologies become increasingly pervasive, speech is emerging as one
of the main input modalities on mobile devices and in intelligent personal
assistants~\cite{McGrawPrabhavalkarAlvarezEtAl16}.
In such applications, speech recognition performance can be improved
significantly by incorporating information about the speaker's context into the
recognition process~\cite{GoogleDQTwiddle2015}.
Examples of such context include the dialog state (e.g., we might want ``stop"
or ``cancel" to be more likely when an alarm is ringing), the speaker's location
(which might make nearby restaurants or locations more
likely)~\cite{ScheinerWilliamsAleksic16}, as well as personalized information
about the user such as her contacts or song
playlists~\cite{AleksicAllauzenElsonEtAl15}.

There has been growing interest recently in building sequence-to-sequence models
for automatic speech recognition (ASR), which directly output words,
word-pieces~\cite{SchusterNakajima12}, or graphemes given an input speech
utterance.
Such models implicitly subsume the components of a traditional ASR system - the
acoustic model (AM), the pronunciation model (PM), and the language model (LM) -
into a single neural network which is jointly trained to optimize
log-likelihood or task-specific objectives such as the expected word error
rate (WER)~\cite{PrabhavalkarSainathWuEtAl18}.
Representative examples of this approach include connectionist temporal
classification (CTC)~\cite{GravesFernandezGomezEtAl06} with word output
targets~\cite{SoltauLiaoSak16}, the recurrent neural network transducer
(RNN-T)~\cite{Graves2012, RaoSakPrabhavalkar17}, and the ``Listen, Attend, and
Spell" (LAS) encoder-decoder architecture~\cite{Chan15,
BahdanauChorowskiSerdyukEtAl16}.
In recent work, we have shown that such approaches can outperform a
state-of-the-art conventional ASR system when trained on $12,500$ hours of
transcribed speech utterances~\cite{CC18}.

In the present work, we consider techniques for incorporating contextual
information dynamically into the recognition process.
In traditional ASR systems, one of the dominant paradigms for incorporating such
information
%into the recognition process
involves the use of an independently-trained
on-the-fly (OTF) rescoring framework which dynamically adjusts the LM weights of a
small number of n-grams relevant to the particular recognition
context~\cite{GoogleDQTwiddle2015}.
%Thus, one of the main benefits of such models is that they do not require the
%development of expertly curated pronunciation lexica, and instead implicitly
%learn to map acoustics to word sequences from the training data, which greatly
%simplifies the process of developing an ASR system.
%However, this simplicity comes at a cost: since the model learns an implicit
%language model from the training data, it is challenging to modify the language
%model probabilities separately from the rest of the model.
Extending such techniques to sequence-to-sequence models is important for
improving system performance, and is an active area of research.
In this context, previous works have examined the inclusion of a separate LM
component into the recognition process through either shallow
fusion~\cite{KannanWuNguyenEtAl18}, or cold fusion~\cite{SriramJunSateeshEtAl17}
which can bias the recognition process towards a task-specific LM.
A shallow fusion approach was also directly used to contextualize LAS
in~\cite{WilliamsKannanAleksicEtAl18} where output probabilities were modified
using a special weighted finite state transducer (WFST) constructed from the
speaker's context, and was shown to be effective in improving
performance.

The use of an external independently-trained LM for OTF rescoring, as in
previous approaches, goes against the benefits derived from the joint
optimization of the components of a sequence-to-sequence model.
Therefore, in this work, we propose Contextual-LAS (CLAS),
a novel, all-neural mechanism which can leverage contextual information --
provided as a list of contextual phrases -- to improve recognition performance.
Our technique consists of first embedding each phrase, represented as a sequence of graphemes, into a fixed-dimensional
representation, and then employing an attention
mechanism~\cite{BahdanauChoBengio15} to summarize the available context at each
step of the model's output predictions.
Our approach can be considered to be a generalization of the technique proposed
in~\cite{Ryan17} in the context of streaming keyword spotting, by allowing for a
variable number of contextual phrases during inference.
The proposed method does not require that the particular context information be
available at training time, and crucially, unlike previous works
~\cite{WilliamsKannanAleksicEtAl18, GoogleDQTwiddle2015}, the method does not
require careful tuning of rescoring weights, while still being able to
incorporate out-of-vocabulary (OOV) terms.
In experimental evaluations, we find that CLAS -- which trains the
contextualization components jointly with the rest of the model -- significantly
outperforms online
rescoring techniques when handling hundreds of context phrases, and is
comparable to these techniques when handling thousands of phrases.

The organization of the rest of this paper is as follows.
In Section~\ref{sec:las} we describe the standard LAS model, and the standard
contextualization approach in Section~\ref{sec:otf-rescoring}.
We present the proposed modifications to the LAS model in order to obtain the
CLAS model in Section~\ref{sec:model}.
We describe our experimental setup and discuss results in
Sections~\ref{sec:expriments} and~\ref{sec:results}, respectively, before
concluding in Section~\ref{sec:conclusion}.

%% file: las.tex
\subsection{The LAS model}\label{sec:las}
%\subsection{LAS model}
% Adapted from:
% https://cs.corp.google.com/piper///depot/google3/experimental/users/tsainath/papers/icassp2018/ronw-multilingual-las/model.tex?type=cs

We now  briefly describe the LAS model. For more details see~\cite{Chan15,CC18}.
The LAS model outputs a probability distribution over sequences of output labels, $\vec{y}$, (graphemes, in this work)
conditioned on a sequence of input audio frames, $\vec{x}$ (log-mel features, in this
work): $P(\vec{y} | \vec{x})$.

\begin{figure}[h]
  \vspace{-2mm} \centering
  \includegraphics[width=\linewidth]{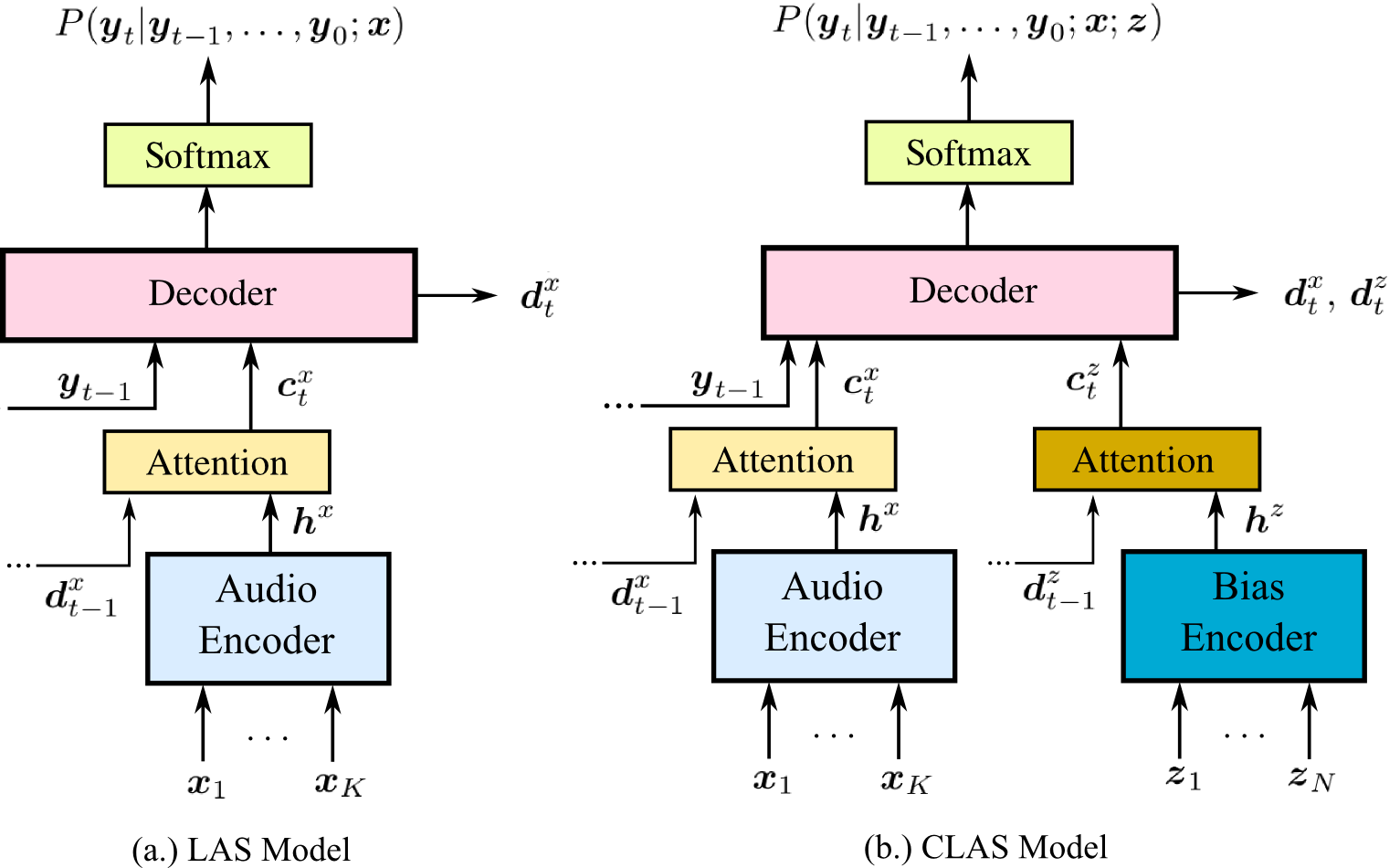}
  \vspace{-7mm}
  \caption{{A schematic representation of the models used in this work. }}
  \label{fig:models}
  \vspace{-0.1in}
\end{figure}

The model consists of three modules:
an \emph{encoder}, \emph{decoder} and \emph{attention network}, which are
trained jointly to predict a sequence of graphemes from a sequence of acoustic
feature frames (Figure \ref{fig:models}a).

%\vspace{-0.07in}
%\subsubsection{Encoder}
The encoder is comprised of a stacked recurrent neural network
(RNN)~\cite{Hochreiter97, Schuster97} (unidirectional, in this work) that reads acoustic features, $\vec{x} = (\vec{x}_1, \dots, \vec{x}_K)$, and
outputs a sequence of high-level features (hidden states), $\vec{h}^x$ =
($\vec{h}^x_1, \dots, \vec{h}^x_K$).
The encoder is similar to the acoustic model in an ASR system.

%\subsubsection{Decoder}
The decoder is a stacked unidirectional RNN that computes the probability of a sequence of
output tokens (characters, in this work) $\vec{y} = (y_1, \dots, y_T)$ as follows:
\begin{equation}
P(\vec{y}|\vec{x}) = P(\vec{y}|\vec{h}^x) = \prod_{t=1}^{T}P(y_{t}|\vec{h}^x, y_0, y_1, \dots, y_{t-1}).
\end{equation}

The conditional dependence on the encoder state vectors, $\vec{h}^x$, is modeled
using a context vector $\vec{c}_{t} = \vec{c}^x_{t}$, which is computed using
Multi-Head-attention~\cite{Vaswani17,CC18} as a function of the current decoder
hidden state, $\vec{d}_t$, and the full encoder state sequence, $\vec{h}^x$.

%The hidden state of the decoder, $\vec{d}_{t}$, captures the previous
%character context $\vec{y_{< t}}$.
The hidden state of the decoder, $\vec{d}_{t}$, which captures the previous
character context $\vec{y_{< t}}$, is given by:
\begin{equation}
\vec{d}_{t} =
\text{RNN}(\tilde{\vec{y}}_{t-1}, \vec{d}_{t-1}, \vec{c}_{t-1})
\end{equation}
where
$\vec{d}_{t-1}$ is the previous hidden state of the decoder, and
$\vec{\tilde{y}}_{t-1}$ is an embedding vector for $y_{t-1}$.
The posterior distribution of the output at time step $t$ is given by:
\begin{equation}
P(y_{t}|\vec{h}^x, \vec{y_{< t}}) = \text{softmax}(\vec{W_\text{s}}[\vec{c}_{t}; \vec{d}_{t}] + \vec{b_\text{s}}),
\end{equation}
where $\vec{W_\text{s}}$ and $\vec{b_\text{s}}$ are again learnable parameters,
and $[\vec{c}_{t}; \vec{d}_{t}]$ represents the concatenation of the two
vectors.
%
%\noindent
The model is trained to minimize the discriminative loss:
\begin{equation}
  L_\text{LAS} = -\log P(\vec{y}|\vec{x})
  \label{eq:las_logprob}
\end{equation}

%% file: standard_context.tex
\subsection{On-the-fly Rescoring \label{sec:otf-rescoring}}

On-the-fly rescoring (similar to~\cite{recency-twiddler}) serves as one of our baseline approaches.
Specifically, we assume that a set of word-level biasing phrases are known ahead
of time, and compile them into a weighted finite state transducer
(WFST)~\cite{MohriPereiraRiley02}.
This word-level WFST, $G$, is then left-composed with a ``speller'' FST, $S$, which
transduces a sequence of graphemes/word-pieces into the corresponding word.
Following the procedure used by~\cite{BahdanauChorowskiSerdyukEtAl16} for a
general language model, we obtain the contextual LM, $C = min(det(S \circ G))$.
The scores from the contextualized LM, $P_C(\vec{y})$, can then be incorporated
into the decoding criterion, by augmenting the standard log-likelihood term with
a scaled contribution from the contextualized LM:
\begin{equation}
\vec{y}^* = \underset{\vec{y}}{\arg\max} \log P(\vec{y}|\vec{x}) + \lambda \log P_{C}(\vec{y})
\label{eq:context}
\end{equation}
\noindent where, $\lambda$ is a tunable hyperparameter controlling how much the
contextual LM influences the overall model score during beam search.

\begin{figure}[h]
  \vspace{-3mm} \centering
  \includegraphics[width=7cm]{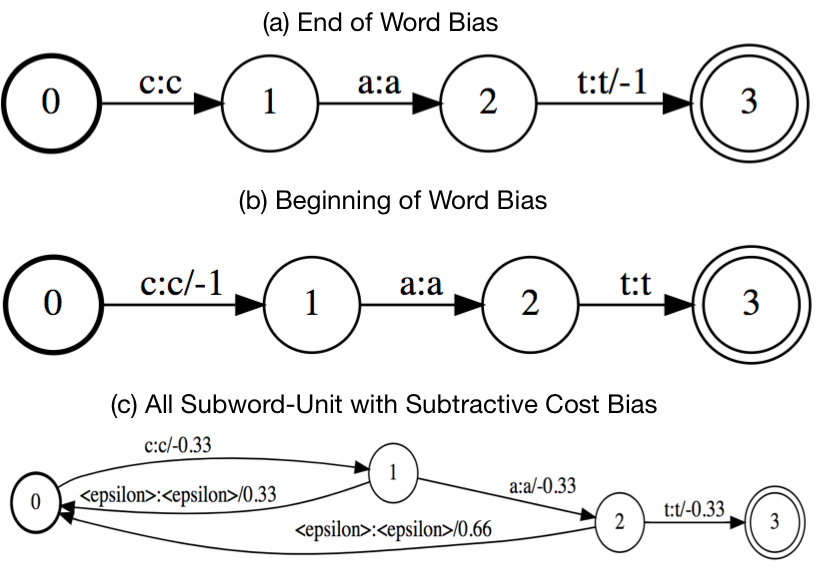}
  \vspace{-3mm}
   \caption{{Different techniques for applying subword-unit scores. Note the costs (i.e, $-1$) are model parameters, and are tuned during inference.}}
  \label{fig:biasing-fst}
  \vspace{-0.1in}
\end{figure}

Note that in~\cite{recency-twiddler}, no weight pushing was applied.
Consequently, the overall score in Equation~\ref{eq:context} is only applied at word
boundaries. This is shown in Figure \ref{fig:biasing-fst}(a). Thus, this technique cannot improve performance if the relevant word
does not first appear on the beam.
Furthermore, we observe that while this approach works reasonably well when the
number of contextual phrases is small (e.g., \texttt{yes}, \texttt{no},
\texttt{cancel})~\cite{recency-twiddler}, it does not work well when the list of
contextual phrases contains many proper nouns (e.g., song names or contacts).
If weight pushing is used, similarly to~\cite{BahdanauChorowskiSerdyukEtAl16}, the
score will only be applied to the beginning subword unit of each word as shown in Figure \ref{fig:biasing-fst}b, which might cause over-biasing problems as we might artificially boost words early on.
%However, weight pushing will apply a score to only one element of the beam, and this could also remove elements from the beam.
Therefore, we explore pushing weights to each subword unit of the word, illustrated in Figure \ref{fig:biasing-fst}c. To avoid artificially giving weight to prefixes which are boosted early on but do not match the entire phrase, we also include a subtractive cost, as indicated by the negative weights in the figure. 
We compare all three approaches in the results section.

%% file: model.tex
\section{Contextual LAS (CLAS)}\label{sec:model}

We will now introduce the Contextualized LAS (CLAS) model which uses additional
context through a list of provided \emph{bias phrases}, $\vec{z}$, thus effectively modeling $P(\vec{y}|\vec{x},
\vec{z})$.
The individual elements in $\vec{z}$ represent phrases such as personalized contact names, song
lists, etc., which are relevant to the particular recognition context.

\subsection{Architecture} \label{sec:clas-description}

We now describe the modification made to the standard LAS model (Figure \ref{fig:models}a)
in order to obtain the CLAS model (Figure \ref{fig:models}b).
The main difference between the two models is the
presence of an additional \emph{bias-encoder} with a corresponding attention
mechanism. These components are described below.

In order to contextualize the model, we assume that the model has access to
%First, we provide the model with
a list of additional sequences of \emph{bias-phrases}, denoted as
$\vec{z} = {\vec{z}_1, \dots, \vec{z}_N}$.
The purpose of the bias phrases is to bias the model towards outputting
particular phrases.
However, not all bias phrases are necessarily relevant given the current
utterance, and it is up to the model to determine which phrases (if any) might
be relevant and to use these to modify the target distribution $P(y_t | \vec{h}^x, \vec{y_{< t}})$.

We augment LAS with a \emph{bias-encoder} which embeds the bias-phrases
into a set of vectors
$\vec{h}^z = \{\vec{h}^z_0, \vec{h}^z_1, \dots, \vec{h}^z_N\}$
(we use superscript $z$ to distinguish bias-attention variables from audio-related variables).
$\vec{h}^z_i$ is an embedding of $\vec{z}_i$ if $i > 0$.
Since the bias phrases may not be relevant for the current utterance, we include
an additional learnable vector, $\vec{h}^z_0 = \vec{h}^z_\text{nb}$, that
corresponds to the the \emph{no-bias} option, that is not using any of the bias
phrases to produce the output.
This option enables the model to backoff to a ``bias-less" decoding strategy
when none of the bias-phrases matches the audio, and allows the model to ignore
the bias phrases altogether.
%Having this option enables the model to fall back
%on a bias-less decoding strategy when none of the bias-phrases matches the audio
%and it is better to ignore the context altogether.
The bias-encoder is a multilayer long short-term memory network
(LSTM)~\cite{Hochreiter97}; the embedding, $\vec{h}^z_i$, is obtained by feeding
the bias-encoder with the sequence of embeddings of subwords in $\vec{z}_i$
(i.e., the same grapheme or word-piece units used by the decoder) and using the last
state of the LSTM as the embedding of the entire phrase~\cite{sutskever:nips14}.

Attention is then computed over $\vec{h}^z$, using a separate attention
mechanism from the one used for the audio-encoder.
A secondary context vector $\vec{c}_t^z$ is computed using the decoder state $\vec{d}_t$.
This context vector summarizes $\vec{z}$ at time step $t$:
\begin{align}
u^z_{it} &= \vec{v}^{z \top} \tanh(\vec{W_h^z}\vec{h}^z_i + \vec{W_d^z}\vec{d}_t + \vec{b^z_\text{a}}) \\
\vec{\alpha}^z_{t} & = \text{softmax}(\vec{u}^z_t) \qquad \vec{c}^z_t = \sum_{i=0}^{N} \alpha^z_{it}\vec{h}^z_{i} \label{eq:bias-attention}
\end{align}

The LAS context vector, which feeds into the decoder, $\vec{c}_t$ is then modified by setting
$\vec{c}_t = \left[\vec{c}_t^x; \vec{c}_t^z\right]$, the concatenation
of context vectors obtained with respect to $\vec{x}$ and $\vec{z}$.
The other components of the CLAS model (i.e., decoder and audio-encoder) are identical
to the corresponding components in the standard LAS model.

\begin{figure}
  \centering
  \includegraphics[width=\linewidth]{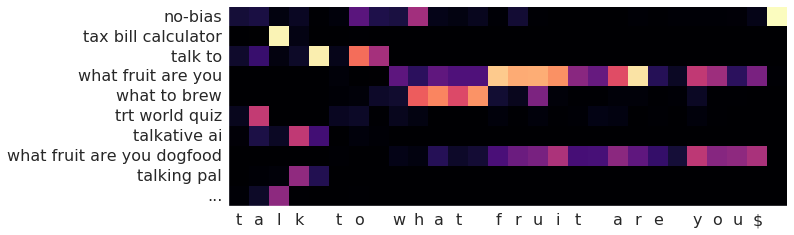}
  %\vspace{-7mm}
  \caption{{Example of \emph{bias-attention-probabilites}, $\vec{\alpha}^z_{t}$, for an utterance
  with reference text \texttt{"talk to what fruit are you"}.
  The utterance was decoded with 3,250 bias phrases, we show here the most active ones.
  Brighter colors denote values closer to 1, while darker colors indicate
  values closer to 0.
  The x-axis is decoding step $t$, and the strings on the left are the most active bias phrases.
  Here we use \texttt{\$} do denote the \texttt{</bias>} token (see
  section~\ref{sec:clas-training}).
  }}
  \label{fig:bias-attn}
  \vspace{-7.2mm}
\end{figure}

It is worth noting that CLAS explicitly models the probability of seeing a particular bias phrase given
the audio and previous outputs:
\begin{equation}
\vec{\alpha}^z_{t} = P(\vec{z_t}|\vec{d}_t) = P(\vec{z_t}|\vec{x} ; \vec{y}_{< t}),
\end{equation}
We refer to $\vec{\alpha}^z_{t}$ as \emph{bias-attention-probability} and an
example of it is presented in Figure~\ref{fig:bias-attn}.

% \subsubsection{Using more sources of context}
% \TS{i'd remove this section as its not used in the paper.}
% So far we have described CLAS with audio and language-biasing sources of context. It
% is easy to add more probabilistic inputs in the same manner, (e.g. georgraphical information, application type etc.)
% by adding additional attention containers and all the resulting context vectors.
% In this work however, we plan to focus only on two sources: audio and language-biasing.

\subsection{Training}
\label{sec:clas-training}
The CLAS model is trained to minimize the loss:
\begin{equation}
L_\text{CLAS} = -\log P(\vec{y}|\vec{x}, \vec{z})
\end{equation}
\noindent where, the bias list, $\vec{z}$, is randomly generated at run time
during training for each training batch.
This is done to allow flexibility in inference, as the model does not make any
assumption about what bias phrases will be used during inference.
The training bias-phrase list is randomly created from the reference
transcripts associated with the utterances in the training batch.
The bias list creation process takes a list of reference transcripts,
$r_1, \dots, r_{N_\text{batch}}$, corresponding to the audio in a training
batch, and randomly selects a list, $\vec{z}$, of n-gram phrases that appear as
substrings in some of the reference transcripts.

To exercise the `no-bias' option, which means that $\vec{z}$ does not match
some of the utterances in the batch, we exclude each reference with probability
$P_\text{keep}$ from the creation process.
When a reference is discarded we still keep the utterance in the batch, but do
not extract any bias phrases from its transcript.
If we set $P_\text{keep} = 0$ no bias phrases will be presented to the training
batch, and $P_\text{keep} = 1$ means that each utterance in the batch will have
at least one matching bias phrase.

Next, $k$ word $n$-grams are randomly selected from each kept reference, where $k$ is
picked uniformly from $[1,N_\text{phrases}]$ and $n$ is picked uniformly from
$[1, N_\text{order}]$.

$P_\text{keep}, N_\text{phrases}$ and $N_\text{order}$ are hyperparameters of the training
process. For example, if we set, $P_\text{keep}=1.0, N_\text{phrases}=1, N_\text{order}=1$, one
unigram will be selected from each reference transcript.
Other choices will be discussed in the experimental section.

Once a set $\vec{z}$ is (randomly) selected, we proceed by computing the
intersection of $\vec{z}$ with each reference transcript $r$. Every time a match
is found a special \texttt{</bias>} symbol is inserted after the match. For
example, if the reference transcript is \texttt{play a song.} and the matching
bias phrase is \texttt{play}, the target sequence will be modified to \texttt{play</bias> a song.}
The purpose of \texttt{</bias>} is to introduce a training error which can be corrected only by
considering the correct bias phrase~\cite{Ryan17}. In other words, to be able to predict \texttt{</bias>}
the model has to attend to the correct bias phrase, thus ensuring that the
\emph{bias-encoder} will receive updates during training.

\subsection{Inference}

During inference, the user provides the system with a sequence of audio feature
vectors, $\vec{x}$, and a set of context sequences, $\vec{z}$, possibly never seen
in training.
Using the bias-encoder, $\vec{z}$ is embedded into $\vec{h}^z$. This embedding can take place before
audio streaming begins.
The audio frames, $\vec{x}$, are then fed into the audio encoder, and the decoder is run as in
standard LAS to produce N-best hypotheses using beam-search decoding~\cite{sutskever:nips14}.

\subsection{Bias-Conditioning \label{sec:conditioning}}

When thousands of phrases are presented to CLAS, retrieving a meaningful bias context vector becomes challenging, since it is the weighted sum
of many different bias-embeddings, and might be far from any context vector seen in training.
Bias-Conditioning attempts to alleviate this problem. Here we assume that during inference
the model is provided with both a list of bias phrases,
$\vec{z} = {\vec{z}_1, \dots, \vec{z}_N}$, as well as a list of \emph{bias
prefixes},
$\vec{p} = {\vec{p}_1, \dots, \vec{p}_N}$.
With this technique a bias phrase $\vec{z}_i$ is ``enabled" at step $t$ only when $\vec{p}_i$ was
detected on the partial hypothesis $\vec{y_{< t}}$ (the partially decoded
hypothesis on the beam).
In practice, we do this by updating the bias-attention-probabilities in
Equation~\ref{eq:bias-attention} by setting:
\begin{align}
\vec{m}_{it} =
\begin{cases}
  0 & \text{if $\vec{p_i} \subseteq$} \vec{y_{< t}} \\
  \infty & \text{otherwise} \\
\end{cases}
\end{align}
\begin{equation}
\vec{\alpha}^z_{t} = \text{softmax}(\vec{u}^z_t - \vec{m_t})
\end{equation}
\noindent where, $\subseteq$ is string inclusion.
The list of bias prefixes can be constructed arbitrarily.
For example, we might want to condition the bias-phrase \texttt{the cat sat} on
the bias-prefix \texttt{the cat}. In this case we will compute an embedding for \texttt{the cat sat}
but ``enable" it only once \texttt{the cat} is detected in $\vec{y_{< t}}$.

A good choice of prefixes will minimize the number of phrases sharing the same
prefix, so the bias-attention is not ``overloaded", while at the same time, not
spliting each phrase into too many segments, to allow distinctive bias embeddings.
This may be achived by an algorithm which starts from empty prefixes ($\vec{p_i} = \epsilon$) and iterativly extends each prefix
by one word (from $\vec{z}_i$) as long as the same prefix is not shared by too many phrases.
In the Section~\ref{sec:clas-cond} we discuss a rule-based prefix construction, and leave full algorithmic treatment as future work.

%% file: experiments.tex
\section{Experiments}\label{sec:expriments}

\subsection{Experimental Setup}

Our training setup is similar to \cite{CC18}, though our
experiments focus on graphemes and our model architecture is smaller. 
Our experiments are conducted on a $\sim$25,000 hour training set consisting of
33 million English utterances.
The training utterances are anonymized and hand-transcribed, and are
representative of Google's voice search traffic.
This data set is augmented by artificially corrupting clean utterances using a
room simulator, adding varying degrees of noise and reverberation such that the
overall SNR is between 0dB and 30dB, with an average SNR of
12dB~\cite{KimMisraChinEtAl17}.
The noise sources are from YouTube and daily life noisy environmental
recordings.

The models evaluated in this section are trained on $8\times8$ Tensor Processing Units (TPU)
slices with global batch size of 4,096. Each training core operates on a shard-size
of 32 utterences in each training step. From this shard, bias phrases are randomized and thus each shard sees a maximum of 32 bias phrases during training.

We use 80-dimensional log-mel acoustic features computed every 10ms over a 25ms window.
Following~\cite{CC18} we stack 3 consecutive frames and stride the stacked
frames by a factor of 3.
This downsampling enables us to use a simpler encoder architecture than \cite{Chan15}.

The encoder's architecture consists of 10 unidirectional LSTM layers, each with $256$ nodes.
The encoder-attention is computed over $512$ dimensions, using $4$ attention heads.
The bias-encoder consists of a single LSTM layer with $512$ nodes and the bias-attention is computed over $512$ dimensions.
Finally, the decoder consists of 4 LSTM layers with $256$ nodes. In total, the model has about 58 million trainable parameters.
Our model is implemented using TensorFlow~\cite{AbadiAgarwalBarhamEtAl15}.

In all our experiments we set $P_\text{keep}=0.5$ to promote robustness to the `no-bias' case.
We set $N_\text{phrases}=1$ and $N_\text{order}=4$.
This leads to a bias list with expected size of $17$ (half of the shard size, plus one for `no-bias').

% To make the model robust to larger bias lists we also augment the
% list with addtional 128 word n-grams selected randomly from the training set.

% We test our model in two use cases of interest.
% In the first case we assume a short list of bias phrases.
% The phrases can be provide manually by a developer based on expected usage.
%
% In the second case we assume a longer list
% that is obtained using a development set. We assume the set is not large enough
% to build a language model, but is sufficient to detect salient terms that will
% help in biasing.

\subsection{Test sets}

We test our model on a number of test sets, which we describe below.
A summary of the biasing setup of each of the test sets is given in
Table~\ref{table:test-sets}.

\begin{table}[h!]
  \vspace{-3mm}
  \centerline{
    \begin{tabular}{ |l|p{1.6cm}|p{1.7cm}|p{1.4cm}| }
    %\begin{tabular}{ |l|c|c|c|c|c| }
      \hline
      Test Set      & Number of utterances & Average bias list size & Bias OOV rate \\
      \hline
      \emph{Voice Search}  & 14k & -    & -     \\
      \emph{Dictation}     & 15k & -    & -     \\
      \emph{Songs}         & 15k & 303  & 3.5\% \\
      \emph{Contacts}      & 15k & 75   & 5.2\% \\
      \emph{Talk-To}       & 4k  & 3,255 & 5.6\% \\
      \hline
    \end{tabular}
  }
\caption{\label{table:test-sets} {\it Details of evaluated test sets. The `bias
OOV rate' measures the fraction of unique words appearing in the bias lists
which are not seen in the training data.}}
\vspace{-0.1in}
\end{table}

The \emph{Voice Search} test set contains voice search queries which are about 3 seconds long.
The \emph{Dictation} test set is contains longer utterances, such as dictations of text messages.
Both \emph{Voice Search} and \emph{Dictation} are in matched conditions to portions
of the training data,
and are not used to test biasing but rather the performance of the model in a bias-free setting.

Each of the remaining test sets: \emph{Songs}, \emph{Contacts}, and
\emph{Talk-To}, contain utterances with a distinct list of contextual phrases which
vary from four phrases upto more than three thousand, and are not necessarily
identical across utterances.

The \emph{Songs}, \emph{Contacts} and \emph{Talk-To} test sets are artificially generated using a \emph{text-to-speech} (TTS) engine.
We use \emph{Parallel WaveNet} \cite{oord2017parallel} as our TTS engine and
corrupt the produced samples with noise similarly to the way we corrupt
the training data.

The \emph{Songs} test set contains requests to play music (e.g. \texttt{play rihanna music}) with a bias set that contains popular
american song and artist names.
The \emph{Contacts} test set contains call requests (e.g. \texttt{call demetri
mobile}) with a bias set that contains an arbitrary list of contact names.

The \emph{Talk-To} test set contains requests to talk with one of many chatbots
(e.g. \texttt{talk to trivia game}). We note that the list of available chatbots
is rather large compared to previous sets. See Table~\ref{table:test-sets} for more details.
%\TS{we should put this statement later??} In Figure~\ref{fig:bias-attn} we present the bias-attention probabilities on one utterance from that set.

%% file: results.tex
\section{Results}\label{sec:results}

In this section, we present the performance of CLAS across a variety of test sets.

\subsection{CLAS without bias phrases}\label{sec:results-nb}

First, to check if our biasing components hurt decoding in cases where no bias phrases
are present, we compare our model to a similar `vanilla' LAS system in table~\ref{table:no-bias-results}.
We note that the CLAS model is trained with random bias phrases, but evaluated
with an empty list of phrases during inference (i.e., only `no-bias' is
presented at inference time), we denote this model by CLAS-NB.
Somewhat surprisingly, we observe that CLAS-NB does better than LAS, and conclude that CLAS can be used even without any biasing phrases.
Therefore, in the experiments that follow, to get accurate comparison, instead of comparing to LAS directly we
use CLAS-NB as proxy for a LAS baseline.

\begin{table} [h!]
  %\vspace{-5mm}
   \centerline{
    \begin{tabular}{ |c|c|c| }
      \hline
      Test Set & LAS WER (\%) & CLAS-NB WER (\%) \\
      \hline
      % baseline model: VSOnlineGraphemeJF
      \emph{Voice Search}  & 6.9 & 6.4 \\
      \emph{Dictation}     & 5.5 & 4.5 \\
      \hline
    \end{tabular}
  }
\caption{\label{table:no-bias-results} {\it LAS vs CLAS where no bias phrases are provided.}}
\vspace{-0.1in}
\end{table}

\subsection{On-the-fly (OTF) Rescoring with LAS Baseline}

Table~\ref {table:fst-bias-results} compares different OTF rescoring variants,
which differ in how weights are assigned to subword units
as outlined in Section~\ref{sec:otf-rescoring}. We only report numbers for the
\emph{Songs} test set; similar trends were observed on the other test sets, which are
omitted in the interest of brevity.
The table indicates that if we bias at the end of the word, as done in \cite{recency-twiddler},
we get very little improvement over the no-bias baseline.
While a small improvement comes from biasing at the beginning of each word \cite{BahdanauChorowskiSerdyukEtAl16},
the best system biases each subword unit, which helps to keep the word on the beam.
All subsequent experiments with OTF rescoring will thus bias each subword unit.

\begin{table}[h!]
  %\vspace{-5mm}
   \centerline{
    \begin{tabular}{ |c|c| }
      \hline
      Method & Songs \\
      \hline
      % baseline model: VSOnlineGraphemeJF
      \emph{No Bias (LAS)}  & 20.9 \\
      \emph{LAS + End of Word Bias~\cite{WilliamsKannanAleksicEtAl18}} &  19 \\
      \emph{LAS + Beginning of Word Bias} & 16.5 \\
      \emph{LAS + Every Subword Unit w/ Subtractive Cost Bias} & 9.4 \\
      \hline
    \end{tabular}
  }
\caption{\label{table:fst-bias-results} {\it LAS with different ways of contextual biasing.}}
\vspace{-0.1in}
\end{table}

\subsection{CLAS with bias phrases}

\subsubsection{Comparison of Biasing Approaches}

\begin{table}[h!]
  \vspace{-3mm}
  \centerline{
    %\begin{tabular}{ |l|p{1.5cm}|p{1.5cm}|p{1.5cm}| }
    \begin{tabular}{ |l|c|c|c| }
      \hline
      Test Set          & LAS & LAS + OTF & CLAS \\
      \hline
      \emph{Songs}      & 18.7 & 9.4  & \textbf{6.9}  \\
      \emph{Contacts}   & 28   & 17.7    & \textbf{7.9}  \\
      \emph{Talk-To}    & 10.8 & \textbf{5.2} & 14  \\
      \hline
    \end{tabular}
  }
\caption{\label{table:blas_vs_baselines} {\it WER for CLAS and basline approaches}}
\vspace{-0.1in}
\end{table}

We compare CLAS to two baseline approaches in Table~\ref{table:blas_vs_baselines}:
(1) A LAS baseline, using CLAS-NB as explained in Section~\ref{sec:results-nb}, (2) LAS + OTF rescoring as described in Section~\ref{sec:otf-rescoring}, with $\lambda$
estimated on the same test sets.
We find that on sets that have hundreds of biasing phrases with high rate of OOVs (Songs, Contacts),
CLAS performs significantly better compared to traditional approaches \emph{without requiring any additional hyperparameter tuning}.
However, CLAS degrades on the \emph{Talk-To} set which has thousands of phrases.
This scalability issue will be addressed with bias-conditioning (see Section~\ref{sec:clas-cond}).

\subsubsection{CLAS with varying number of bias phrases}

\begin{figure}[h]
  \vspace{-3mm} \centering
  \includegraphics[width=6cm]{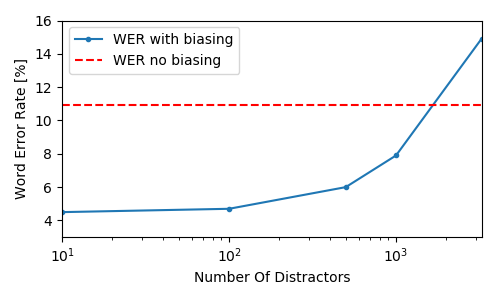}
  \vspace{-3mm}
   \caption{{WER for the \emph{Talk-To} set with a CLAS model (without rescoring or conditioning).
  }}
  \label{fig:talk-to-scale}
  \vspace{-0.1in}
\end{figure}

To better understand the CLAS failure with \emph{Talk-To} we evaluate CLAS while
restricting to phrases that appear in the reference transcript plus $N$ \emph{distractors} (phrases which are not present in the transcript)
  chosen randomly from the complete bias list. The results are presented in Figure~\ref{fig:talk-to-scale}.
We observe gradual degradation in WER as a function of number of distractors.

We hypothesise that the reason for this behavior is that with a large number of
bias phrases, correlations start to appear between their embeddings.
For example, the embeddings of \texttt{talking pal} and \texttt{talkative ai} have a correlation (normalized inner product) of 0.6,
while the average correlation is 0.2.
%We address the scaling issue with bias-conditioning in the next section.

\subsubsection{Overcoming the scaling problem: CLAS with Bias-conditioning (Cond) and OTF rescoring}\label{sec:clas-cond}

\begin{table}[h!]
  \vspace{-1mm}
  \centerline{
    \begin{tabular}{ |l|p{1.3cm}|p{1.3cm}|p{1.3cm}|p{1.3cm}| }
    %\begin{tabular}{ |l|c|c|c|c }
      \hline
      Test Set          & CLAS & CLAS + OTF & CLAS + Cond & CLAS + Cond + OTF \\
      \hline
      \emph{Songs}      & 6.9  & \textbf{5.7} & - & -  \\
      \emph{Contacts}   & 7.9  & \textbf{7.5} & - & -  \\
      \emph{Talk-To}    & 14   & 9.0 & 7.4 & \textbf{5.6}  \\
      \hline
    \end{tabular}
  }
\caption{\label{table:blas_rescoring} {\it WER for biasing test set. The results show the benefit for CLAS from both Conditioning and OTF rescoring.}}
\vspace{-0.1in}
\end{table}

Next, we try to combine CLAS with bias-conditioning (Section~\ref{sec:conditioning}).
Since \emph{Talk-to} has the largest number of biasing phrases, we test scalability of CLAS by applying bias-conditioning to this set only, with
prefixes constructed in a rule-based manner:
First we create a prefix from \texttt{talk to} + the next word, (e.g. the phrase \texttt{talk to pharmacy flashcards},
would be split into a prefix $p_i=\texttt{talk to pharmacy}$ and a suffix $z_i = \texttt{flashcards}$).
In addition we found it useful to condition the first word after \texttt{talk to} on its first letter (e.g. \texttt{pharmacy} will be conditioned on \texttt{talk to p}).
This construction restricts the number of phrases sharing the same prefix to 225 (vs. 3255) while increasing the overall number of bias phrase segments
by only 10\%.

In Table~\ref{table:blas_rescoring} with show our bias-conditioning and OTF-rescoring (Section~\ref{sec:otf-rescoring}) CLAS results.
CLAS benefits from either approach, as well as from their combination.
Indeed, conditioning allows us to scale to a large number of phrases without any degradation.

%% file: conclusion.tex
\section{Conclusions}\label{sec:conclusion}
In this work we presented CLAS, a novel, all-neural, end-to-end contexualized
ASR model, which incorporates contextual information by embedding full context
phrases.
In experimental evaluations, we demonstrated how the proposed model outperforms
standard shallow-fusion biasing techniques on several test sets.
%We hypothesize that the main issue with shallow-fusion is the direct change of the model output,
%which may lead to discrepancy with the model's internal state.
%Instead, CLAS directly augments the model inputs with context.
We investigated CLAS's ability to handle a large set of context phrases, and suggested a conditioning method to further imporve
its quality.
Our future work includes scaling CLAS to tens of thousands of bias phrases.